\documentclass{aip-cp}

\usepackage[numbers]{natbib}
\usepackage{rotating}
\usepackage{graphicx}
\usepackage[english]{babel}
\usepackage[T1]{fontenc}
\usepackage[utf8]{inputenc}

\newcommand{\fermilat}{Fermi-LAT}
\newcommand{\hess}{H.E.S.S.}
\newcommand{\pg}{PG~1553+113}
\newcommand{\pks}{PKS~2155-304}
\newcommand{\cena}{Cen~A}

\begin{document}

\title{Hadronic Modeling of TeV AGN: Gammas and Neutrinos}

\author[aff1]{M. Cerruti\corref{cor1}}
\author[aff2]{A. Zech}
\author[aff3]{G. Emery}
\author[aff3]{D. Guarin}

\affil[aff1]{Sorbonne Universités, UPMC, Université Paris Diderot, Sorbonne Paris Cité, CNRS, LPNHE, 4 place Jussieu, F-75252, Paris Cedex 5, France}
\affil[aff2]{LUTH, Observatoire de Paris, CNRS, Université Paris Diderot, PSL Research University, 5 Place Jules Janssen, 92190 Meudon,
France}
\affil[aff3]{UPMC, 4 place Jussieu, F-75252, Paris Cedex 5, France}
\corresp[cor1]{mcerruti@lpnhe.in2p3.fr}

\maketitle

\begin{abstract}
Blazar emission models are usually divided into two big families, leptonic and hadronic, according to the particles which are responsible for the gamma-ray component of their spectral energy distributions. Even though leptonic models have been successful in explaining the gamma-ray emission from most blazars, hadronic models still present an intriguing alternative. They have the unique advantage to link the emission of photons, neutrinos, and cosmic rays from the astrophysical source.
We have developed a stationary one-zone lepto-hadronic code to model the emission from blazars in both leptonic and hadronic scenarios. In this contribution we focus on hadronic modeling of a few selected TeV blazars of the BL Lac type and the radio galaxy Centaurus A detected by Cherenkov telescopes. We study in particular their associated neutrino emission, for different hadronic scenarios, and how it compares to the sensitivity of current neutrino detectors.
\end{abstract}

\section{Blazar hadronic models}
Blazars are a class of active galactic nuclei (AGN) characterized by non-thermal emission from radio to $\gamma$-rays, rapid variability (down to the minute time-scale), and polarized emission in visible wavelengths. They are considered as radio-loud AGN whose relativistic jet points in the direction of the Earth, and the photon emission is associated with the non-thermal population of particles in the jet. Blazars represent the large majority of AGN detected at $\gamma$-rays, both at MeV-GeV energies with Fermi-LAT, and at TeV energies with Cherenkov telescopes such as H.E.S.S., MAGIC and VERITAS. The spectral energy distribution (SED) of blazars is always comprised of two non-thermal bumps: the low-energy one peaks between infrared and X-rays, while the high-energy one peaks at $\gamma$-rays, from MeV to TeV. Blazar can be classified according to the energy of the first SED component: flat-spectrum-radio-quasars (FSRQs) peak, in general, in infrared, while BL Lacertae objects show different peak energies from infrared (low-frequency-peaked objects, LBL) to UV/X-rays (high-frequency-peaked objects, HBL). At TeV energies, the extragalactic sky is largely dominated by HBLs.\\

The origin of the first SED component in blazars is generally associated with synchrotron emission by a population of electrons (and positrons) in the relativistic jet. Polarization measurements in radio support this scenario. The origin of the high-energy SED component is, on the other hand, more disputed. In leptonic scenarios, it is associated with an inverse Compton scattering between the electrons/positrons in the jet and their own synchrotron emission (synchrotron-self-Compton, SSC) or an external photon field (external-inverse-Compton, EIC). In hadronic models it is associated with synchrotron emission by protons, or by secondary particles produced in proton-photon interactions.\\

Neutrinos are naturally produced in hadronic models, due to the pion production in p-$\gamma$ interactions, and their subsequent decay into leptons and neutrinos ($\nu_\mu$ and $\nu_e$). In recent years, the first detections of astrophysical neutrinos with IceCube \citep{icecubefirst}, has opened a new window on the sky, and renewed interest in blazar hadronic models. So far, no point-like source of neutrinos has been detected, and the current detections are compatible with an isotropic flux. In order to compare the predictions of blazar hadronic models with the IceCube sensitivity, we make use of the sensitivity curves for a detection of a point-like source at the $5\sigma$ confidence level in four years of IceCube observations, as provided in \citep{icecube4years}.\\

In \cite{leha}, we presented a new stationary lepto-hadronic code to simulate blazar emission. The code computes all the relevant leptonic and hadronic emission processes, and allows the user to simulate leptonic or hadronic scenarios in a consistent framework. We recall here briefly the main characteristics of the code:
\begin{itemize}
\item the leptonic part is an evolution of the stationary leptonic code described in \cite{Kata}; 
\item proton-photon interactions are computed using the Monte-Carlo code SOPHIA \cite{sophia}; 
\item Bethe-Heitler pair-production is computed analytically following \cite{Kelner08};
\item synchrotron-supported pair cascades are computed iteratively
\end{itemize}

Current blazar observations have not been able to determine unequivocally which emission process and type of particles are responsible for the blazar emission. Given that no neutrinos are produced in a purely leptonic model, a smoking gun for hadronic processes would be the detection of neutrino emission from a blazar. Such a measurement would constrain blazar emission models and at the same time establish AGN as hadron accelerators, and potentially solve the open question of the origin of ultra-high-energy cosmic rays (UHECRs; with $\textrm{E} > 10^{18}$ eV). The neutrino emission is computed directly by the SOPHIA code, and we can extract the expected neutrino fluxes for a given hadronic model which fits the blazar photon SED. The neutrino emission varies significantly within the acceptable model parameter space, even for a given set of data from a specific source. The goal of this work is to investigate and present the expected neutrino emission for different successful blazar hadronic models for different TeV blazars. In Section 1 we present the results for modeling of the TeV HBL \pks ; in Section 2 we study the radio-galaxy Centaurus A, modeled as a mis-aligned blazar; in Section 3 we focus on the distant HBL \pg.

\section{A bright TeV HBL : \pks\ }
\label{sec1}
 \pks\ (z=0.116) is one of the brightest extragalactic TeV sources, together with Mrk 421 and Mrk 501, showing regularly a TeV flux at the level of a few 10\% of the Crab nebula or more. In 2006, the \hess\ telescopes detected a bright TeV flare from this source \citep{2155flare} which, due to its exceptional brightness and variability down to the minute timescale, stands today as one as the most exceptional TeV flares ever seen. Given that hadronic models are, in general, disfavoured for rapid flares, mainly due to the longer acceleration and cooling time-scales for protons, we focus instead on a hadronic modeling of the low-state SED taken from \citep{21552009}.\\

\begin{figure}
\includegraphics[width=400pt]{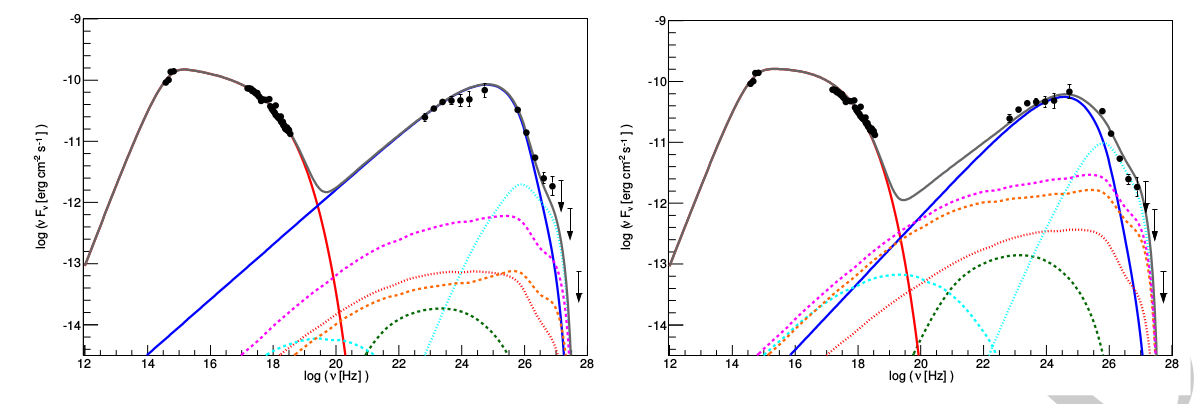}
\caption{Hadronic modelling of \pks. The left shows a model obtained assuming a magnetic field strength $\log B [G] = 0.3$, while the right plot shows a model obtained assuming $\log B [G] = 1.9$ and a smaller radius. Synchrotron emission from primary electrons is shown with the red solid line; synchrotron emission from primary protons is shown with the blue solid line; synchrotron-self-Compton emission is shown with the dashed green line; synchrotron emission by muons is shown with the dotted light-blue line; synchrotron emission by cascades triggered by $\pi^0$ decay, $pi^\pm$ decay and Bethe-Heitler pair-production re shown with the dashed violet line, dashed red line and dotted red line, respectively. Taken from \citep{leha2} \label{PKS2155fig}}
\end{figure}

Following \citep{leha}, we reduce the number of free parameters by: i) assuming co-acceleration of protons and electrons, and thus a same value for the index of their injected energy distribution (assumed to be a power-law with exponential cut-off); ii) assuming simple synchrotron cooling for the electrons; iii) fixing the maximum proton energy as a function of the size $R$ and the magnetic field $B$ (via an equation between the acceleration and cooling time-scales); iv) fixing the Doppler factor $\delta=30$. Under these assumptions we can systematically study the parameter space, to find the family of solutions which can reproduce the SED. The first result is that, for the case of \pks, all hadronic solutions are characterized by proton-synchrotron emission dominating the MeV-GeV part of the SED, while in the TeV regime synchrotron emission from secondary particles (muons and electrons/positrons) produced in p-$\gamma$ interactions can be important. The strength of this secondary component varies as a function of the model parameters, as can be seen in Figure \ref{PKS2155fig}, in which we reproduce, as an example among many, two distinct hadronic models of \pks. The potential detectability of this ''cascade bump'' with CTA is presented in details in \citep{lehaicrc, leha2}.\\

\begin{figure}
\includegraphics[width=200pt]{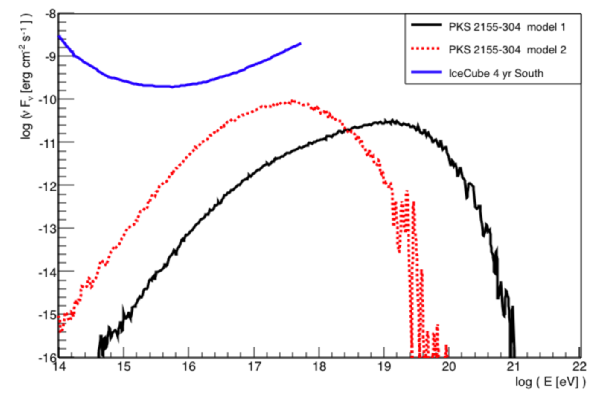}
\caption{Neutrino emission expected in the hadronic emission scenario shown in Fig. \ref{PKS2155fig}: the black line and the red line represent the neutrino spectrum associated with the model in Fig. \ref{PKS2155fig}, right, and left, respectively. The blue line represents the expected sensitivity of IceCube to point-like sources in four years of observations. Adapted from \citep{leha2}. \label{PKS2155neutrinofig}}
\end{figure}

The corresponding neutrino emission is shown in Figure \ref{PKS2155neutrinofig}. The first important information is that the neutrino emission varies significantly between two hadronic models which have similar photon emission. In particular, the peak energy of the neutrino emission strongly depends on the maximum proton energy. The second important information is that the expected neutrino flux is still much lower than the sensitivity of IceCube for point-like sources and, at least during a flux state similar to the low-state observed in 2009, no neutrino emission from \pks\ is expected to be observed with IceCube, based on its four-year sensitivity curve.\\


\newpage
\section{A TeV Radio-Galaxy : Centaurus A }
\label{sec2}
\begin{figure}
\includegraphics[width=460pt]{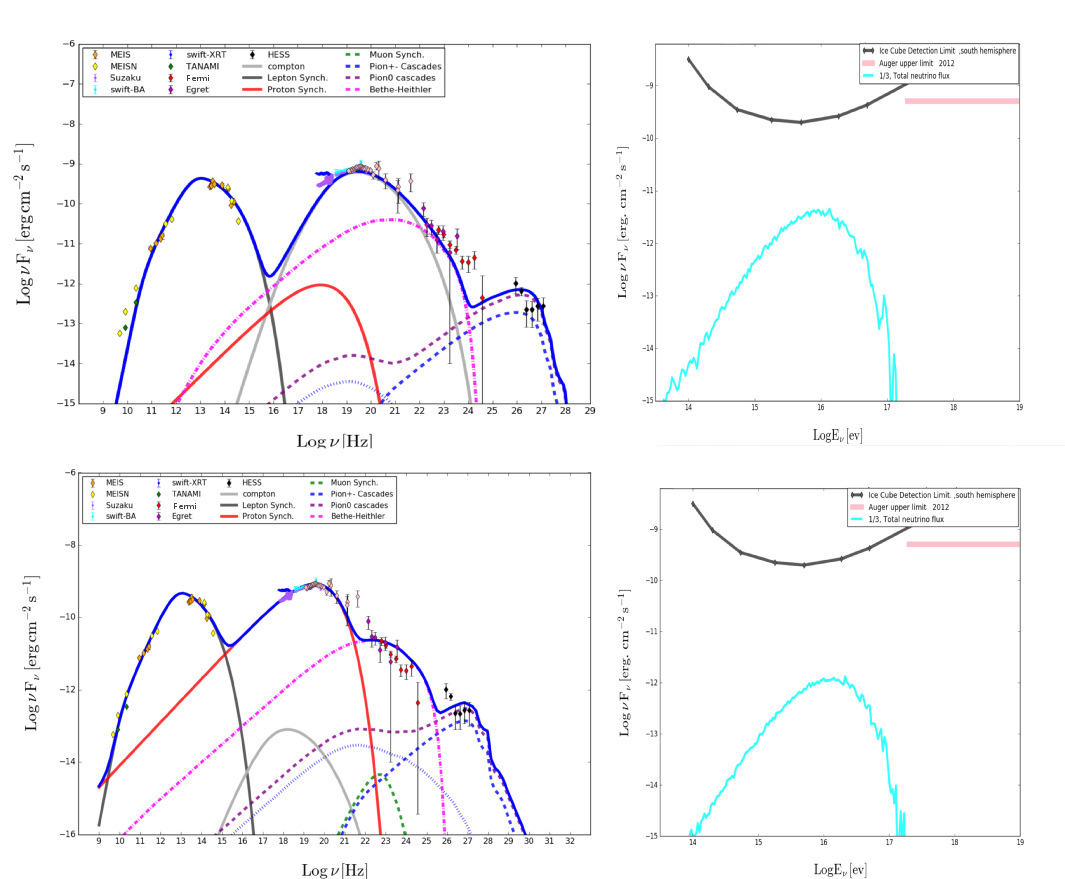}
 \caption{Hadronic modeling of the radio-galaxy \cena. The top plot shows a solution dominated by SSC emission in X-rays, while the bottom solutions is dominated by proton-synchrotron emission in X-rays. The color code is different from Fig. \ref{PKS2155fig}, and detailed in the inset. Data have been taken from \citep{fermicena}. On the right plot, we show the corresponding neutrino emission ($\nu_e + \nu_\mu$), compared to the sensitivity of IceCube to point-like sources in four years of observations. \label{CenAfig}}
\end{figure}

Centaurus A is a very well known radio-galaxy, located at $z=0.00183$. Discovered as source of TeV photons by H.E.S.S. \citep{cenahess}, it is one of the few radio-galaxies detected with Cherenkov telescopes. Its SED is significantly different from the ones of typical HBLs, such as \pks: a first peak, associated to synchrotron emission by electrons, is seen in infrared, while the second SED bump, peaking in X-rays, extends up to TeV. Recently, \fermilat\ observations have shown evidence of a spectral break in the GeV regime, suggesting the emergence of a new component which connects smoothly to the H.E.S.S. data \citep{saha13, brown16}. Such an additional component, difficult to explain with standard leptonic models \citep{petrocena}, can be naturally produced in hadronic models, due to secondary particles coming from p-$\gamma$ interactions. In Fig. \ref{CenAfig} we show two different hadronic one-zone models for Cen A. In the first one (Fig. \ref{CenAfig}, top) the model is basically an SSC scenario up to the MeV regime, but at higher energies the synchrotron emission from secondary pairs coming from p-$\gamma$ interactions with the primary electron-synchrotron radiation become important. In the second model (Fig. \ref{CenAfig}, bottom), the X-ray peak is associated with proton-synchrotron emission, while the emission observed with Fermi-LAT and \hess\ is again due to secondary pairs. In particular, emission from cascades produced in the Bethe-Heitler pair-production are expected to dominate the Fermi-LAT spectrum in both scenarios. In the left part of Fig. \ref{CenAfig} we show the associated neutrino emission for our models: in both cases the neutrino flux from Cen A is expected to peak at tens of PeV, and is still significantly lower (by about two orders of magnitude) than the sensitivity of IceCube to point-like sources.\\

\begin{figure}
\includegraphics[width=260pt]{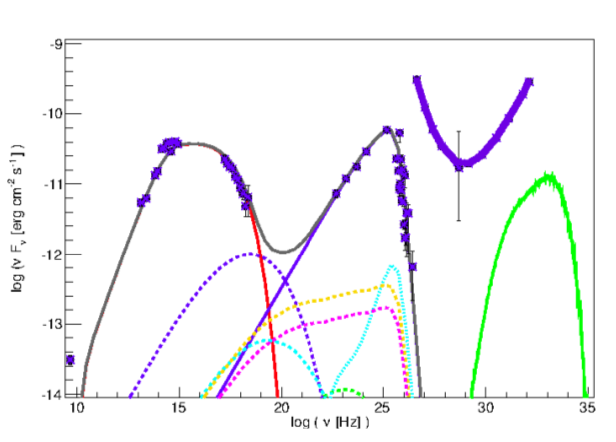}
\includegraphics[width=260pt]{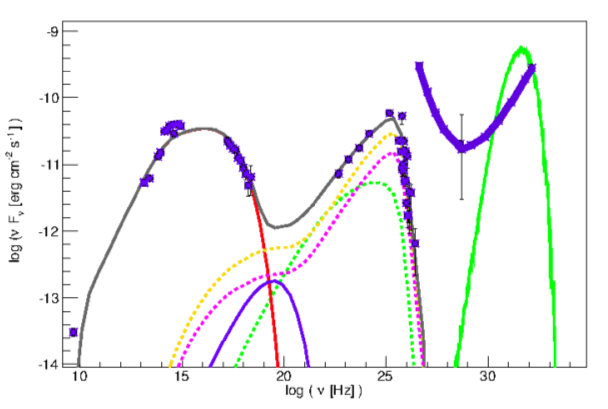}
\caption{Hadronic modeling of \pg. In this case we represent on the same plot the photon emission and the neutrino emission. The color code for the radiative components of the hadronic modeling is the same as in Fig. \ref{PKS2155fig}. The total neutrino emission ($\nu_e + \nu_\mu$) is shown with the green solid line. The IceCube sensitivity to point-like sources of neutrinos in four years of observations is represented with the violet solid line. The data point at an energy of $0.2$ PeV is NU-17 (see text).  \label{PG1553fig}}
\end{figure}

\section{A distant TeV HBL : \pg\ }
\label{sec3}
Located at a redshift $0.395 \leq z \leq  0.58$ \citep{Danforth10}, \pg\ is one of the most distant persistent emitters of TeV photons, regularly monitored by H.E.S.S., MAGIC and VERITAS. Among the first list of astrophysical neutrinos detected with IceCUBE \citep{icecubefirst}, the event number NU-17, with an energy of $200 \pm 27$ TeV, was reconstructed at only 8.9$^\circ$ from \pg, making this blazar a possible counterpart \citep{Padovani14, Petro15}. In this case as well, we investigated possible hadronic solutions of the photon emission from \pg, comparing the associated neutrino emission with the IceCUBE measurement and its sensitivity to point-like sources. In the following we fix z=0.4 as source redshift. The left plot of Fig. \ref{PG1553fig} shows a typical proton-synchrotron solution of the SED (built using the ASDC SED builder tool), in which the full H.E.S.S. spectrum is associated to protons, with only marginal contribution from secondary particles. In this case, similarly to the modeling of \pks\ presented above, no neutrino emission is expected to be detected with IceCube (again, for this specific flux state in photons; given the highly variable nature of blazars, conclusions from modeling of flaring states may be different). At the energy of the detected IceCube neutrino NU-17 (0.2 PeV), the model emission in neutrinos is about three orders of magnitudes lower than the data. \\
In the right plot of Fig. \ref{PG1553fig} we present instead a lepto-hadronic solution, in which the $\gamma$-ray emission is associated with synchrotron emission from secondary leptons produced in p-$\gamma$ interactions and SSC radiation. This scenario, as pointed out by \citep{Petro15}, is much more favorable for neutrino emission, although the expected flux is at energies much higher than NU-17. For the solution presented here (which represents only one possible realisation of the hadronic modeling) the neutrino emission is indeed higher than the sensitivity of IceCUBE to point-like sources, and in this case we would expect a much higher number of detected neutrinos coming from a position compatible with \pg. In such a case, the non-detection by IceCube of neutrino point-like sources can be used indeed to constrain, and potentially exclude, at least part of the broad parameter space of blazar hadronic models, strengthening constraints based on the power of the emitting region.\\

\section{Conclusions}
The emergence of neutrino astronomy is triggering renewed interest into hadronic models for blazars. In this context we investigated the neutrino emission expected from a hadronic modeling of three well known gamma-ray sources: the HBL \pks, the radio-galaxy Cen A, and the distant blazar \pg. For the case of \pks, we have shown that the expected neutrino emission is much lower than the neutrino sensitivity, at least for a low-state in the blazar $\gamma$-ray emission. Detection of a neutrino emission would be on the other hand extremely useful to remove the degeneracy of the hadronic model, constraining the hadronic content of the relativistic jet. For the case of Cen A, hadronic models represent an appealing alternative to explain the complex $\gamma$-ray spectrum of the source, which shows a hardening at GeV energies, with a minimum set of free parameters. In this case as well, no neutrino emission is expected to be detected with IceCube. Finally we investigated the neutrino emission from \pg, which is particularly interesting due to its proximity to one of the IceCube detections. In this case, while pure proton-synchrotron models are not interesting for neutrinos, a mixed lepto-hadronic model predicts indeed an emission in neutrinos even higher than the IceCube sensitivity, which thus can begin to constrain at least part of the parameter space.


\newpage
 \section{Acknowledgements}
The authors wish to thank J. Finke for providing the published multi-wavelength data for the Centaurus A core emission SED.

\nocite{*}
\bibliographystyle{aipnum-cp}%
\bibliography{leha}%

\end{document}